\documentclass[aps,prl,twocolumn,superscriptaddress,amsmath,amssymb]{revtex4}
\usepackage{graphicx}

\begin{document}
\title{Collective Dipole Oscillation of a Spin-Orbit Coupled Bose-Einstein Condensate}
\author{Jin-Yi Zhang}
\author{Si-Cong Ji}
\affiliation{Hefei National Laboratory for Physical Sciences at Microscale and Department of Modern Physics, University of Science and Technology of China, Hefei, Anhui 230027, PR China }
\author{Zhu Chen}
\affiliation{Institute for Advanced Study, Tsinghua University, Beijing, 100084, China}
\author{Long Zhang}
\author{Zhi-Dong Du}
\author{Bo Yan}
\author{Ge-Sheng Pan}
\author{Bo Zhao}
\affiliation{Hefei National Laboratory for Physical Sciences at Microscale and Department of Modern Physics, University of Science and Technology of China, Hefei, Anhui 230027, PR China }
\author{You-Jin Deng}
\email{yjdeng@ustc.edu.cn}
\affiliation{Hefei National Laboratory for Physical Sciences at Microscale and Department of Modern Physics, University of Science and Technology of China, Hefei, Anhui 230027, PR China }
\author{Hui Zhai}
\email{hzhai@mail.tsinghua.edu.cn}
\affiliation{Institute for Advanced Study, Tsinghua University, Beijing, 100084, China}
\author{Shuai Chen}
\email{shuai@ustc.edu.cn}
\affiliation{Hefei National Laboratory for Physical Sciences at Microscale and Department of Modern Physics, University of Science and Technology of China, Hefei, Anhui 230027, PR China }\author{Jian-Wei Pan}
\email{pan@ustc.edu.cn}
\affiliation{Hefei National Laboratory for Physical Sciences at Microscale and Department of Modern Physics, University of Science and Technology of China, Hefei, Anhui 230027, PR China }
\date{\today}

\begin{abstract}
In this letter we present an experimental study of the collective dipole oscillation of a spin-orbit coupled Bose-Einstein condensate in a harmonic trap. 
Dynamics of the center-of-mass dipole oscillation is studied in a broad parameter region, 
as a function of spin-orbit coupling parameters as well as oscillation amplitude.
Anharmonic properties beyond effective-mass approximation are revealed, such as amplitude-dependent frequency
and finite oscillation frequency at place with divergent effective mass.
These anharmonic behaviors agree quantitatively with variational wave-function calculations. 
Moreover, we experimentally demonstrate a unique feature of spin-orbit coupled system predicted by a sum-rule approach, 
stating that spin polarization susceptibility--a static physical quantity--can be measured via dynamics of dipole oscillation.
The divergence of polarization susceptibility is observed at the quantum phase transition that separates 
magnetic nonzero-momentum condensate from nonmagnetic zero-momentum phase.
The good agreement between the experimental and theoretical results provides a bench mark for recently developed theoretical approaches.
\end{abstract}
\maketitle

Many interesting quantum phases can emerge in solid state materials when electrons
are placed in a strong magnetic field or possess strong spin-orbit (SO) coupling,
such as the fractional quantum Hall effect~\cite{FQH} and the topological insulator~\cite{TI}.
In cold atom systems, albeit neutral atoms have neither charges nor SO coupling, the recent exciting experimental
progress demonstrates that artificial gauge potentials can be synthesized in laboratory
by laser control technique \cite{NIST,NIST_elec,SOC,NIST_partial,GFinlattice,JZhang-2011,SOC_Fermi,SOC_MIT}.
Synthetic gauge potential is becoming a powerful tool
for simulating real materials with cold atoms. Moreover, the system of SO coupled bosons does not have an analogy in
conventional condensed matter systems, and can exhibit many novel phases \cite{review} such as
striped superfluid phase \cite{Stripe,Ho} and half vortex phase \cite{Wu,Victor,Hu,Santos}.

Collective excitations play an important role in studying physical
properties of trapped atomic Bose-Einstein condensates (BEC) and degenerate Fermi gases.
Collective dipole oscillation is a center-of-mass motion of all atoms.
For a conventional condensate, the dipole oscillation is trivial: 
the frequency is just the harmonic-trap frequency, independent of oscillation amplitude and interatomic interaction.
This is known as Kohn theorem  \cite{stringari,Kohn}.
For a SO coupled condensate, however, it was found \cite{NIST_elec} that the dipole-oscillation frequency 
deviates from the trap frequency and the experimental data thereby can be explained by effective-mass approximation. 
Recently, much theoretical effort has been taken to understand dynamics of a SO coupled BEC \cite{duine,Hu_mode,Chuan_Wei,Chen,Yongping,Yun}, 
and many predicted unconventional properties remain to be experimentally explored. 
In particular, the so-called sum-rule approach predicts \cite{Yun} a unique feature of SO coupled condensate: 
spin polarization susceptibility--a static physical quantity--can be measured via dynamics of dipole oscillation.

In this letter we experimentally study the collective dipole oscillation of a SO coupled $^{87}$Rb BEC, 
occurring both in momentum and magnetization.  The oscillation frequency is measured along various paths 
in the phase diagram, as a function of SO-coupling parameters and oscillation amplitude.
Anharmonic properties beyond effective-mass approximation are observed, including amplitude-dependent frequency 
and finite oscillation frequency at place with infinite effective mass. 
The experimental data fit well with variational wave-function calculations.
Moreover, following the proposal by the sum-rule approach \cite{Yun}, we deduce the spin polarization susceptibility 
from the amplitude ratio between momentum and magnetization oscillations; 
the results are in good agreement with theoretical calculations. 
In particular, the measured spin polarization susceptibility does diverge 
at the quantum phase transition that separates magnetic nonzero-momentum condensate from nonmagnetic zero-momentum phase.
To the best of our knowledge, this is the first experimental measurement of a divergent spin susceptibility 
at magnetic phase transition in cold atom system.

\begin{figure}
  \includegraphics[height=3.5in, width=3.4in]{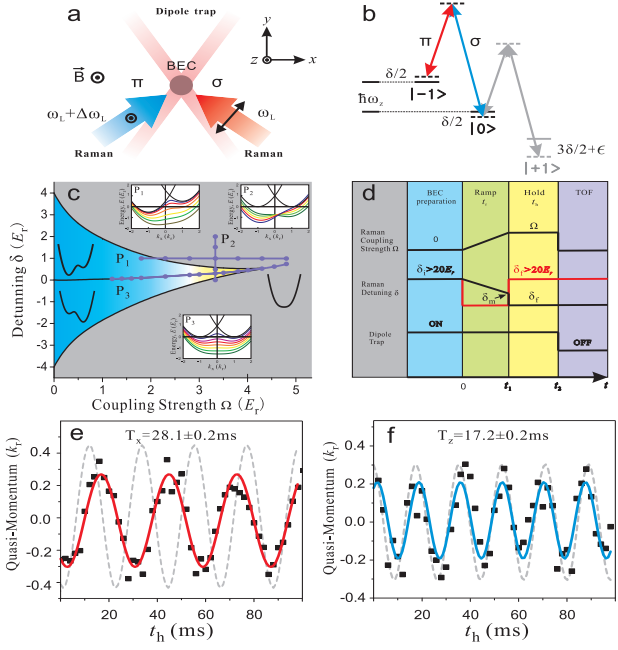}\\
  \caption{ (a) Experimental layout. Field $\mathbf{B}_{\text{bias}}$ is along the $z$ direction
       and the Raman lasers propagate in the $x$-$y$ plane.
   (b) Raman coupling scheme within the $F=1$ manifold.
   (c) Single-particle phase diagram in the $\Omega$-$\delta$ plane.
       The dispersion spectrum $\mathcal{E}(k_{x})$ has two (one) local minima in the blue (grey) regime.
       Experiments are along paths $\text{P}_{1}$, $\text{P}_{2}$,
       and $\text{P}_{3}$, with $\mathcal{E}(k_{x})$ in the insets.
   (d) Experimental time sequence. (e-f) Dipole oscillation with $\Omega=3.3E_{\text{r}}$
          and $\delta=1E_{\text{r}}$ along $\hat{x}$(e) and $\hat{z}$(f) direction.
	  For comparison, oscillations without SO coupling are displayed as the grey dashed curves.}
\label{exp}
\end{figure}

The experimental layout is sketched in Fig.~\ref{exp}(a).
A $^{87}\mathrm{Rb}$ BEC of about $2.5 \times 10^5$ atoms is produced in the crossed optical dipole
trap with wavelength $1070$nm, beam waist 80$\mathrm{\mu m}$, and trap frequency $\omega= 2 \pi \times $ \{45,45,55\}Hz.
A bias magnetic field $\mathbf{B}_{\text{bias}}$ is applied in the $z$ direction.
The BEC is illuminated by a pair of Raman lasers in the $x$-$y$ plane
with beam waist $ 240\mathrm{\mu m}$, relative angle $\theta=105^{\circ}$ and wavelength $\lambda=803.3$nm.
The Raman lasers couple the three internal states of the $F=1$ manifold as shown in Fig.~\ref{exp}(b).
By setting the quadratic Zeeman shift $\epsilon= 3.37 E_{\text{r}}$ with recoil
energy $E_{\text{r}}= k^2_{\text{r}}/(2m)=2\pi\times2.21$kHz ($k_{\text{r}}$ is the recoil momentum),
we effectively suppress state $|m_\text{F}= 1\rangle$
and prepare a spin-$1/2$ system by regarding state $|m_\text{F}=-1 \rangle $  as  $| \uparrow \rangle $
and $|m_\text{F}=0 \rangle $ as $ |\downarrow\rangle $.
This leads to a single-particle Hamiltonian as ($\hbar=1$)
\begin{equation}
   \hat{H}_0=
   {\begin{pmatrix}
     \frac{(k_{x}+k_{\text{r}})^{2}}{2m} -\frac{\delta}{2} & \frac{\Omega}{2} \\
     \frac{\Omega}{2} & \frac{(k_{x}-k_{\text{r}})^{2}}{2m}+\frac{\delta}{2}   \\
        \end{pmatrix}}
 \label{3matrix}
 \end{equation}
where $k_{x}$ is the quasi-momentum, $\Omega$ is the strength of Raman coupling, and $\delta$
is the two-photon Raman detuning.
Diagonalization of Eq.~(\ref{3matrix}) yields two eigenstates for each $k_x$, in which
the spin and the momentum of an atom are coupled.
For small $\Omega$ and $\delta$, energy dispersion $\mathcal{E}(k_{x})$ has two local minima,
and bosons condense in one of the minima for our current experimental conditions.
For large $\Omega$ or $\delta$, $\mathcal{E}(k_{x})$ has only one minimum.
The phase boundary between the double- and the single-minimum region is displayed in Fig.~\ref{exp}(c),
in which the perturbative effect of state $|m_\text{F}= 1\rangle$ has been taken into account 
(as well as in theoretical calculations later).
We shall systematically study the dipole oscillation
along paths $\text{P}_1, \text{P}_2$, and $\text{P}_3$ in Fig.~\ref{exp}(c).

The time sequence for the experiment is shown in Fig.~\ref{exp}(d).
After the BEC is prepared in the trap ($t=0$),
the Raman coupling is adiabatically ramped up from zero to $\Omega$
in a time period $t_{1}$ from 70 to 100 ms, and  $\mathbf{B}_{\text{bias}}$ is slowly ramped,
which adiabatically changes $\delta$ from an initial
value $\delta_{\text{i}}>20E_{\text{r}}$ to an intermediate value $\delta_{\text{m}}$.
At $t_1$, detuning $\delta$ is switched from $\delta_{\text{m}}$ to $\delta_{\text{f}}$ (black line in Fig.~\ref{exp}(d)) in 1ms, 
which is much faster than the oscillation period and slow enough that the BEC remains in the lower eigenstate. 
This process effectively gives the BEC a pulsed momentum, named synthetic electric force in Ref. \cite{NIST_elec}.
Further, by varying $\delta_{\text{m}}$, one excites the dipole oscillation with different amplitudes.
The BEC starts to oscillate at $t_1$ and is held for a holding time $t_{\text{h}} = t_2-t_1$.
At $t_2$ both the Raman lasers and the trap are quickly switched off within 1$\mathrm{\mu s}$.
With the Stern-Gerlach technique, a time-of-flight (TOF) image
is taken after 24ms of free expansion to map out the spin and momentum of the BEC.
A comparison experiment without SO coupling is also carried out
by setting a large detuning $\delta>20E_\text{r}$
instead of turning off the Raman lasers.
Accordingly, the time sequence for $\mathbf{B}_{\text{bias}}$ is modified:
$\mathbf{B}_{\text{bias}}$ is initially set for $\delta = \delta_{\text{f}}$ at $t=0$ and
quickly switched to be $\delta>20E_\text{r}$ at $t_2$ (red line in Fig.~\ref{exp}(d)).

\begin{figure}
  \includegraphics[width=3.0in]{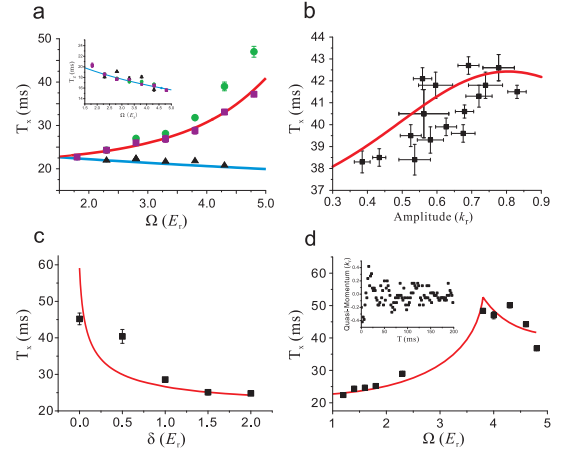}
  \caption{
	 (a) Oscillation period  $T_{x}$ along path $\text{P}_{1}$ ($\delta=1E_{\text{r}}$).
          versus $\Omega$.  Black triangles are without SO coupling,
	       and the blue line denotes the estimated trap frequency.
	       Purple squares are with SO coupling for relatively small (below $0.2k_r$) oscillation amplitude,
           while green circles are for large (about $0.6k_r$) amplitude.
	       The inset is for $T_{z}$ versus $\Omega$.
	  (b) $T_{x}$ versus oscillation amplitude with $(\Omega=4.8, \delta=1.0)E_{\text{r}}$.
	       Black squares with error bars are experimental data.
	       (c) and (d), $T_x$ versus $\Omega$ along path $\text{P}_2$ and path $\text{P}_3$.
	       In all plots, the red lines are the theory curves by solving Eq.~(\ref{rdot}).
	  }
  \label{Frequency}
\end{figure}

{\it Momentum oscillation.}  Figure \ref{exp}(e-f) displays typical momentum oscillations observed in the experiment.
The frequency along $\hat{x}$ is significantly changed by the SO coupling,
while those along $\hat{y}$ and  $\hat{z}$ remain to be the trap frequency (independent of oscillation amplitude).
Figure \ref{Frequency}(a) shows oscillation period $T_x$ versus  $\Omega$ along path $\text{P}_1$
($\delta = 1  E_{\text{r}}$).
As $\Omega$ increases, the deviation from the trap frequency becomes larger.
Moreover, it is found that $T_x$ is amplitude-dependent, as shown in Fig.~\ref{Frequency}(b) for $ \Omega = 4.8  E_{\text{r}}$
and $\delta = 1 E_{\text{r}}$. These two features
clearly demonstrate that the dipole oscillation along the $x$ direction is no longer harmonic.

Similar $\Omega$-dependence of $T_x$ along $P_2$ and $P_3$ are displayed in Fig. \ref{Frequency}(c) and (d).
In the yellow regime of Fig.~\ref{exp}(c), the single-particle spectrum has two nearly degenerate minima 
separated by a barrier as low as trapping energy. 
Macroscopic quantum tunneling between two minima takes place in this regime, and 
the dipole oscillation becomes rather complicated and does not fit a single-frequency oscillation  
(shown in the inset of Fig. \ref{Frequency}(d)). 

This anharmonic behavior can be understood from the equation-of-motion. Consider Hamiltonian
$\hat{H}=\sum_i (H_{0,i}+(1/2) m{\bf \omega}_\alpha^2  r^2_{\alpha i})+ \sum_{i<j}U({\bf r}_i-{\bf r}_j)$, 
where $\alpha$ sums over $x$, $y$ and $z$, $U({\bf r})$ represents two-body interaction. 
Let $\hat{X} : \equiv (1/N) \sum_{i} x_{i}$ be the center-of-mass displacement operator ($N$ be the number of atoms),
the equation-of-motion for $X$ can be derived as $ \dot{\hat{X}}= 1/(N\,m)\sum_i(\hat{k}_{x,i}+k_\text{r}\sigma_{z,i})$ 
and $\ddot{\hat{X}} =-\omega^2_{x} \hat{X}+i\Omega k_\text{r}\sum_i \sigma_{y,i} $.
For either no coupling between spins ($\Omega=0$) or momentum-independent coupling--e.g. coupled by radio-frequency field with $k_{\text{r}}=0$,
the equation-of-motion will close at the second order, which yields harmonic oscillation and the well-known Kohn theorem~\cite{stringari,Kohn}.
For both nonzero $\Omega$ and $k_{\text{r}}$, however, the equation-of-motion cannot close at any finite order. 
This results in anharmonic dipole oscillation.

For a quantitative calculation of the oscillation frequency,
we shall apply a variational wave-function approach \cite{Zoller}. 
We first assume that the condensate stays in the lower eigen-branch during the entire oscillation \cite{Chen}.
We further ignore spin-dependent interaction because 
the spin-dependent interaction energy is about 0.46$\%$ of the total energy for  the $F=1$ manifold of $^{87}$Rb atoms \cite{c_2}
and the aspect ratio of condensate in our experiments is far away from ``mode resonance" \cite{Chen}. 
With these simplifications, the dipole oscillation is described by \cite{Chen}
\begin{equation}
 \dot{k}_x = -\omega^2_{x} x \; , \hspace{6mm}
\dot{x}   = \partial \mathcal{E}(k_x)/ \partial k_x \; , \label{rdot}
\end{equation}
As shown in Fig.~\ref{Frequency}, our experimental data agree well with calculations based on Eq.~(\ref{rdot}).
In particular, for ($\Omega=4 E_{\text{r}}$, $\delta=0$) where effective-mass approximation breaks down because of divergent effective mass,
Fig.~\ref{Frequency} shows that the oscillation frequency remains finite. 

\begin{figure}
  \includegraphics[width=3.0in]{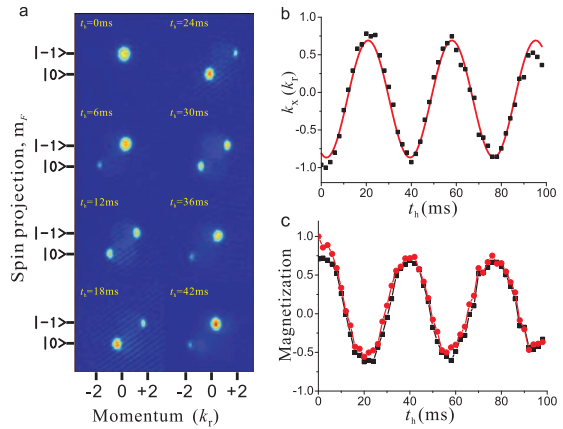}
  \caption{
  Magnetization oscillation for $\delta=1E_{\text{r}}$ and $\Omega=4.8E_{\text{r}}$.
          (a), Spin-resolved TOF images for various holding times $t_h$.
          (b), Quasi-momentum $k_x$ versus $t_{\text{h}}$.
          (c), Polarization $\mathcal{M}$ versus $t_{\text{h}}$.
	       Red circles are directly measured, while
               black squares are deduced from (b) (see text for details).
	  }
  \label{mag}
\end{figure}

{\it Magnetization Oscillation.}
The Stern-Gerlach TOF images in Fig.~\ref{mag}(a) show that during the dipole oscillation, 
the spin population also oscillates.
Figure~\ref{mag}(b) and (c) display the oscillation of quasi-momentum $k_x$ and 
of polarization $\mathcal{M}=(n_{\uparrow}-n_{\downarrow})/(n_{\uparrow}+n_{\downarrow})$, respectively;
it can be seen that their frequency is exactly identical.
To provide an intuitive picture, we assume that during the entire oscillation,
the BEC remains at the lower eigenstate branch of Hamiltonian~(\ref{3matrix}), and the spin configuration
adiabatically follows the center-of-mass motion.
The eigenstate wave function of the lower branch $\psi = f_\uparrow(k_x) |\uparrow \rangle
+ f_\downarrow(k_x) |\downarrow \rangle $ can be obtained by diagonalizing Eq.~(\ref{3matrix}),
which yields  $\mathcal{M}(k_x) = |f_\uparrow(k_x)|^2-|f_\downarrow(k_x)|^2$. Hence, the magnetization changes with $k_x$.
The function $\mathcal{M}(t)$ can be then obtained by using the experimentally measured $k_x (t)$ (Fig.~\ref{mag}(b)).
The results are shown as black squares in Fig.~\ref{mag}(c), and agree very well with the directly measured data (red circles). 
The magnetization oscillation reflects the locking between spin and momentum in Hamiltonian (\ref{3matrix}), 
and further provides a direct justification for the assumption in the variational wave-function approach.

The magnetization oscillation can be understood from the absence of Galilean invariance. 
Consider a BEC that moves with velocity $v$ along $\hat{x}$, in the co-moving frame the single-particle Hamiltonian 
acquires an additional term $ vk_x$. In a conventional BEC, this term can be gauged away by a gauge transformation $\psi \rightarrow e^{i mvx} \psi$. 
In our system, however, such a procedure will introduce a velocity-dependent Zeeman-energy term  $-m v k_{\text{r}}\sigma_z$. 
Hence, once the condensate moves, an oscillation of magnetization $\mathcal{M}$ has to be induced.

\begin{figure}
  \includegraphics[width=3.4in]{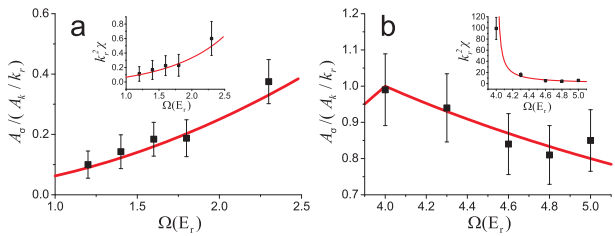}\\
  \caption{Amplitude ratio of spin and momentum oscillation $A_{\sigma}/(A_{k}/k_{\text{r}})$ versus $\Omega$, 
  for magnetic phase (a) and nonmagnetic phase (b).
  The inset is for the spin polarization susceptibility $E_{\text{r}} \, \chi$ deduced from this ratio
  via Eq. (\ref{relation}). The red solid line is from the solution of Eq. (\ref{eq:chi}).
	  }
  \label{susceptibility}
\end{figure}

{\it Spin Polarization Susceptibility and Quantum Phase Transition}. 
A unique feature of SO coupled condensate is that spin polarization susceptibility can be deduced from the amplitude ratio between 
momentum and magnetization oscillations~\cite{Yun}.  Here we focus on $\delta=0$. 
For $\Omega<4 E_{\text{r}}$, bosons condense in one of the double minima, which spontaneously breaks the time-reversal symmetry 
and has non-zero magnetization $\langle\sigma_z\rangle$. 
For $\Omega>4 E_{\text{r}}$, bosons condense in zero-momentum state with zero magnetization. 
Thus, there is a phase transition from magnetic condensate to nonmagnetic condensate as $\Omega$ varies \cite{Yongping,Yun}. 
The spin polarization susceptibility $\chi$ can be expressed as \cite{Yun} 
\begin{equation}
  \chi = \left\{ 
         \begin{array}{lcl} 
           \frac{\Omega^2/2E_{\text{r}}}{16E^2_{\text{r}}-\Omega^2} \hspace{5mm} & \mbox{for} & \Omega < 4 E_\text{r}  \\
	   \frac{2}{\;\; \Omega-4E_{\text{r}}\;\;}                  \hspace{5mm} & \mbox{for} & \Omega > 4 E_\text{r} 
	  \end{array} 
	  \right.
\label{eq:chi}
\end{equation}
Equation~(\ref{eq:chi}) predicts  that $\chi$ diverges at the phase transition point $\Omega=4E_{\text{r}}$.
It is further proposed \cite{Yun} that $\chi$ can be measured from the amplitude ratio between spin and momentum oscillations 
via 
\begin{equation}
\frac{A_\sigma}{A_k/k_{\text{r}}}=\frac{E_{\text{r}}\chi}{1+E_{\text{r}}\chi} \; . \label{relation}
\end{equation}
We experimentally measure ratio $A_\sigma/(A_k/k_{\text{r}})$ and compare the data to the theoretical prediction by Eq. (\ref{eq:chi}).
Figure \ref{susceptibility}(a) and (b) shows the results for magnetic phase with $\Omega<2.5 E_{\text{r}}$ and
for nonmagnetic phase $\Omega>4E_{\text{r}}$, respectively. 
Note that for the reason discussed above, no experimental data are available for 2.5$E_{\text{r}}<\Omega<4E_{\text{r}}$. 
We further deduce susceptibility $\chi$ via Eq. (\ref{relation}) from the experimental ratio $A_\sigma/(A_k/k_{\text{r}})$, 
shown in the inset of Fig. \ref{susceptibility}. 
The excellent agreement with the theoretical result confirms the unique feature of SO coupled 
condensate that spin susceptibility can be measured via dipole oscillation.
In particular, one finds that as $\Omega \rightarrow 4E_{\text{r}}+0^+$, $\chi$ does display divergent behavior,
giving a strong evidence of quantum phase transition.

In summary, we have experimentally demonstrated non-trivial properties of the dipole oscillation for a SO coupled BEC. 
From the experimentally measured dynamics of dipole oscillation, we further display the divergent behavior of 
spin polarization susceptibility--a static physical quantity--at the quantum phase transition.
Beside being a direct experimental observation of unconventional dynamics in a SO coupled condensate, 
our quantitative results also provide a bench mark for various recently developed theoretical approaches. 
It is expected that further study of dynamic behavior would provide a powerful tool
in probing novel phases of the SO coupled BEC, such as stripe superfluid phase \cite{Stripe,Ho}.

HZ would like to thank Sandro Stringari and Lev Pitaevskii for valuable discussions.
The measurement of spin polarization susceptibility via Eq.~(\ref{relation}) was suggested by Sandro Stringari.
This work has been supported by the NNSF of China, the CAS,
the National Fundamental Research Program (under Grant No. 2011CB921300, No. 2011CB921500) and NSERC.



\vspace*{-2mm}
\begin{thebibliography}{99}
\bibitem{FQH}
D. C. Tsui, H. L. Stormer, and A. C. Gossard, Phys. Rev. Lett. {\bf 48}, 1559 (1982).

\bibitem{TI}
For a review, see M. Z. Hasan, and C. L. Kane, Rev. Mod. Phys. {\bf 82}, 3045 (2010);
X.-L. Qi, and S.-C. Zhang, Rev. Mod. Phys. {\bf 83}, 1057 (2011).

\bibitem{NIST}
Y.-J. Lin, R. L. Compton, A. R. Perry, W. D. Phillips, J. V. Porto, and I. B. Spielman, Phys. Rev. Lett. \textbf{102}, 130401 (2009);
Y.-J. Lin, R. L. Compton, K. Jim\'{e}nez-Garc\'{\i}a, J. V. Porto, and I. B. Spielman, Nature \textbf{462}, 628 (2009).

\bibitem{NIST_elec}
Y.-J. Lin, R. L. Compton, K. Jim\'{e}nez-Garc\'{\i}a, W. D. Phillips, J. V. Porto, and I. B. Spielman, Nature Physics {\bf 7}, 531 (2011).

\bibitem{SOC}
Y.-J. Lin, K. Jim\'{e}nez-Garc\'{\i}a, and I. B. Spielman, Nature \textbf{471}, 83 (2011).

\bibitem{NIST_partial}
R. A. Williams, L. J. LeBlanc, K. Jim\'{e}nez-Garc\'{\i}a, M. C. Beeler, A. R. Perry, W. D. Phillips, and I. B. Spielman, Science \textbf{335}, 314 (2012).

\bibitem{GFinlattice}
M. Aidelsburger, M. Atala, S. Nascimb\`{e}ne, S. Trotzky, Y.-A. Chen, and I. Bloch, Phys. Rev. Lett. {\bf 107}, 255301 (2011).

\bibitem{JZhang-2011}
Z. Fu, P. Wang, S. Chai, L. Huang , and J. Zhang, Phys. Rev. A  {\bf 84}, 043609 (2011).

\bibitem{SOC_Fermi} 
P. Wang, Z.-Q. Yu, Z. Fu, J. Miao, L. Huang, S. Chai, H. Zhai, and J. Zhang, arXiv:1204.1887.

\bibitem{SOC_MIT} 
L. W. Cheuk, A. T. Sommer, Z. Hadzibabic, T. Yefsah, W. S. Bakr, and M. W. Zwierlein, arXiv:1205.3483.

\bibitem{review}
For a review, see H. Zhai, Int. J. Mod. Phys. B {\bf 26}, 1230001 (2012).

\bibitem{Stripe}
C. Wang, C. Gao, C.-M. Jian, and H. Zhai, Phys. Rev. Lett.
{\bf 105}, 160403 (2010).

\bibitem{Ho}
T.-L. Ho, and S. Zhang, Phys. Rev. Lett. {\bf 107}, 150403 (2011).

\bibitem{Wu}
C.-J. Wu, I. Mondragon-Shem, and X.-F. Zhou, Chin. Phys. Lett. {\bf 28} 097102 (2011).

\bibitem{Victor}
T. D. Stanescu, B. Anderson, and V. Galitski, Phys. Rev. A {\bf 78}, 023616 (2008).

\bibitem{Hu}
H. Hu, B. Ramachandhran, H. Pu, and X.-J. Liu, Phys. Rev. Lett. \textbf{108}, 010402 (2012).

\bibitem{Santos}
S. Sinha, R. Nath, and L. Santos, Phys. Rev. Lett. {\bf 107}, 270401 (2011).

\bibitem{stringari}
S. Stringari, Phys. Rev. Lett. {\bf 77}, 2360 (1996).

\bibitem{Kohn}
W. Kohn, Phys. Rev. {\bf 123}, 1242 (1961);
F. Dalfovo, S. Giorgini, L. P. Pitaevskii, and S. Stringari, Rev. Mod. Phys. \textbf{71}, 463 (1999).

\bibitem{duine}
E. van der Bijl, and R. A. Duine, Phys. Rev. Lett. {\bf 107}, 195302 (2011).

\bibitem{Hu_mode}
B. Ramachandhran, B. Opanchuk, X.-J. Liu, H. Pu, P. D. Drummond, and H. Hu,
Phys. Rev. A \textbf{85}, 023606 (2012).

\bibitem{Chuan_Wei}
Y. Zhang, L. Mao, and C. Zhang, Phys. Rev. Lett. {\bf 108}, 035302 (2012).

\bibitem{Yongping}
Y. Zhang, G. Chan and C. Zhang, arXiv: 1111.4778.

\bibitem{Chen}
Z. Chen and H. Zhai, arXiv: 1204.5121.

\bibitem{Yun}
Y. Li, G. Martone and S. Stringari, arXiv: 1205.6398.

\bibitem{Zoller}
V. M. P\'{e}rez-Garc\'{\i}a, H. Michinel, J. I. Cirac, M. Lewenstein, and P. Zoller, Phys. Rev. Lett.
{\bf 77}, 5320 (1996).

\bibitem{c_2}
A. Widera, F. Gerbier, S. F\"{o}lling, T. Gericke, O. Mandel, and I. Bloch, N. J. Phys. {\bf 8}, 152 (2006).

\end{thebibliography}
\end{document}